\documentclass[twoside,11pt]{article}

\usepackage[preprint]{jmlr2e}

\usepackage{amsmath}
\usepackage{blindtext}
\usepackage{booktabs}
\usepackage{caption, subcaption}
\usepackage[frozencache=true]{minted}
\usepackage{hyperref}
\usepackage{makecell}
\usepackage{lipsum}
\usepackage{csquotes}
\usepackage{tikz}
\usepackage{xcolor}

\hypersetup{colorlinks=true}

\DeclareUnicodeCharacter{3BC}{$\mu$}
\DeclareUnicodeCharacter{22C5}{.}

\newcommand{\inline}[1]{\mintinline{Julia}{#1}}
\definecolor{bg}{rgb}{0.95,0.95,0.95}

\definecolor{blue}{rgb}{0.251,0.388,0.847}
\definecolor{green}{rgb}{0.220,0.596,0.149}
\definecolor{purple}{rgb}{0.584,0.345,0.698}
\definecolor{red}{rgb}{0.796,0.235,0.200}
\definecolor{white}{rgb}{1,1,1}

\usetikzlibrary{math, backgrounds, arrows, shapes, positioning}
\usetikzlibrary{arrows.meta}

\tikzstyle{edge} = [->, line width=1.0pt, {round cap}-{Latex[round, scale=0.7]}, shorten <= 2pt, shorten >= 2pt]
\tikzstyle{node} = [draw=none, align=center, text=white,]
\tikzstyle{roundnode} = [node, circle, minimum size = 0.5cm]
\tikzstyle{rectanglenode} = [node, rectangle, minimum height = 1.4em]
\tikzstyle{roundednode} = [rectanglenode, rounded corners = 0.1cm]

\makeatletter
\newcommand*{\centerfloat}{%
  \parindent \z@
  \leftskip \z@ \@plus 1fil \@minus \textwidth
  \rightskip\leftskip
  \parfillskip \z@skip}
\makeatother

\usepackage{lastpage}
\jmlrheading{26}{2025}{1-\pageref{LastPage}}{}{}{}{Guillaume Dalle and Adrian Hill}

\ShortHeadings{A Common Interface for Automatic Differentiation}{Dalle and Hill}
\firstpageno{1}

\begin{document}

\title{A Common Interface for Automatic Differentiation}

\author{\name Guillaume Dalle \email guillaume.dalle@enpc.fr \\
  \addr LVMT, ENPC, Institut Polytechnique de Paris, Univ Gustave Eiffel, Marne-la-Vallée, France
  \AND
  \name Adrian Hill \email hill@tu-berlin.de \\
  \addr BIFOLD – Berlin Institute for the Foundations of Learning and Data, Berlin, Germany\\
  Machine Learning Group, Technical University of Berlin, Berlin, Germany}

\editor{My editor}

\maketitle

\begin{abstract}
  For scientific machine learning tasks with a lot of custom code, picking the right Automatic Differentiation (AD) system matters.
  Our Julia package \texttt{DifferentiationInterface.jl} provides a common frontend to a dozen AD backends, unlocking easy comparison and modular development.
  In particular, its built-in preparation mechanism leverages the strengths of each backend by amortizing one-time computations.
  This is key to enabling sophisticated features like sparsity handling without putting additional burdens on the user.
\end{abstract}


\section{Motivation}

Automatic Differentiation (AD) is a cornerstone of modern machine learning \citep{baydinAutomaticDifferentiationMachine2018}.
By generating derivatives directly from computer code, AD obviates the need for manual differentiation of complex algorithms.
While this separation of concerns enables quick prototyping, it requires compatibility between each application and one or more AD systems.
For standardized tasks like deep learning, AD is often used as part of an integrated framework, which makes compatibility straightforward.
Python programmers may pick \texttt{TensorFlow} \citep{abadiTensorFlowLargescaleMachine2015}, \texttt{PyTorch} \citep{paszkePyTorchImperativeStyle2019}, or \texttt{JAX} \citep{bradburyJAXComposableTransformations2018},
then write all of their code within the boundaries of that framework, from neural layers to optimization routines.
Yet if the task is not standardized, and if the software stack is not already set in stone (e.g.\ by technical or collaboration constraints), \enquote{shopping around} for the best AD solution can be hugely beneficial.

Indeed, not all AD libraries offer the same mutation support, looping abilities, branching behavior, or sparsity handling.
These variations impact performance when the function to differentiate is not a typical neural network.
A crucial example is scientific machine learning, which seeks to build differentiable models of physical processes.
The modeling code might involve non-vectorizable procedures, in-place memory updates, custom iterations, nested control flow and scalar indexing, which are a struggle for deep learning-oriented frameworks.
When one cannot pick the best AD system \emph{a priori}, perhaps one should write the code first, and compare options later?
In the present paper, we show that this is possible.

Since its inception a decade ago, the Julia language \citep{bezansonJuliaFreshApproach2017} has emerged as a worthy contender for numerical computing and scientific machine learning.
In that community, the freedom to choose between AD systems has been a longstanding goal \citep{sapienzaDifferentiableProgrammingDifferential2024}.
This flexibility is a social necessity, because Julia's AD landscape is less centralized and well-funded than Python's, leaving key packages in the hands of students or academics with temporary positions.
Luckily, it is also within reach, because Julia's built-in arrays and operations are sped up by just-in-time compilation.
As a result, AD does not need to target a very specific subset of accelerated primitives (like \texttt{jax.numpy}): in theory, the whole language lends itself to differentiation, which facilitates separation between AD and its uses.
In practice of course, each AD library imposes a different set of tradeoffs and limitations.

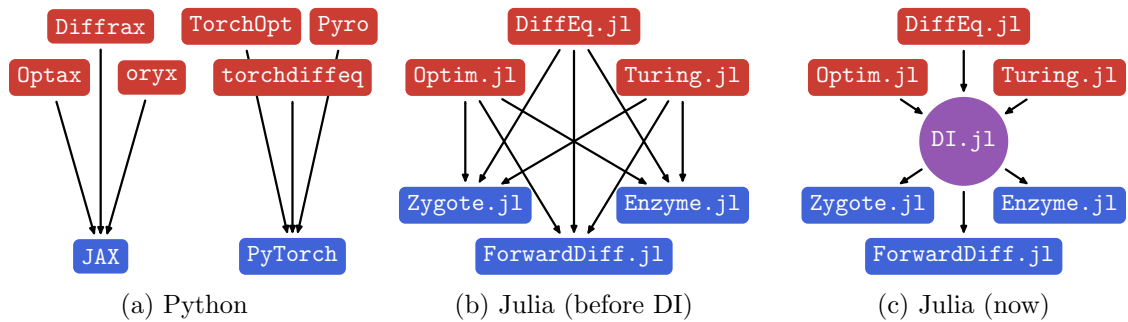
\begin{figure}
  \begin{subfigure}{0.32\textwidth}
    \centering
    \scalebox{0.85}{
      \begin{tikzpicture}
        \node[roundednode, fill=red] (Optax) at (-0.8, 0) {\texttt{Optax}};
        \node[roundednode, fill=red] (Diffrax) at (0, 0.8) {\texttt{Diffrax}};
        \node[roundednode, fill=red] (oryx) at (0.8, 0) {\texttt{oryx}};

        \node[roundednode, fill=red] (TorchOpt) at (2.2, 0.8) {\texttt{TorchOpt}};
        \node[roundednode, fill=red] (torchdiffeq) at (3, 0) {\texttt{torchdiffeq}};
        \node[roundednode, fill=red] (Pyro) at (3.8, 0.8) {\texttt{Pyro}};

        \node[roundednode, fill=blue] (JAX) at (0, -2.8) {\texttt{JAX}};
        \node[roundednode, fill=blue] (PyTorch) at (3, -2.8) {\texttt{PyTorch}};

        \begin{scope}[on background layer]
          \draw[edge] (Optax) to (JAX);
          \draw[edge] (Diffrax) to (JAX);
          \draw[edge] (oryx) to (JAX);

          \draw[edge] (TorchOpt) to (PyTorch);
          \draw[edge] (torchdiffeq) to (PyTorch);
          \draw[edge] (Pyro) to (PyTorch);
        \end{scope}
      \end{tikzpicture}
    }
    \subcaption{Python}
  \end{subfigure}
  \hfill
  \begin{subfigure}{0.32\textwidth}
    \centering
    \scalebox{0.85}{
      \begin{tikzpicture}
        \node[roundednode, fill=red] (Optim) at (-0.2, 0) {\texttt{Optim.jl}};
        \node[roundednode, fill=red] (DiffEq) at (1.5, 0.8) {\texttt{DiffEq.jl}};
        \node[roundednode, fill=red] (Turing) at (3.2, 0) {\texttt{Turing.jl}};

        \node[roundednode, fill=blue] (Zygote) at (-0.2, -2) {\texttt{Zygote.jl}};
        \node[roundednode, fill=blue] (ForwardDiff) at (1.5, -2.8) {\texttt{ForwardDiff.jl}};
        \node[roundednode, fill=blue] (Enzyme) at (3.2, -2) {\texttt{Enzyme.jl}};

        \draw[edge] (Optim) to (Zygote);
        \draw[edge] (Optim) to (ForwardDiff);
        \draw[edge] (Optim) to (Enzyme);

        \draw[edge] (DiffEq) to (Zygote);
        \draw[edge] (DiffEq) to (ForwardDiff);
        \draw[edge] (DiffEq) to (Enzyme);

        \draw[edge] (Turing) to (Zygote);
        \draw[edge] (Turing) to (ForwardDiff);
        \draw[edge] (Turing) to (Enzyme);
      \end{tikzpicture}
    }
    \subcaption{Julia (before DI)}
  \end{subfigure}
  \hfill
  \begin{subfigure}{0.32\textwidth}
    \centering
    \scalebox{0.85}{
      \begin{tikzpicture}
        \node[roundednode, fill=red] (Optim) at (-0, 0) {\texttt{Optim.jl}};
        \node[roundednode, fill=red] (DiffEq) at (1.5, 0.8) {\texttt{DiffEq.jl}};
        \node[roundednode, fill=red] (Turing) at (3, 0) {\texttt{Turing.jl}};

        \node[roundednode, fill=blue] (Zygote) at (-0, -2) {\texttt{Zygote.jl}};
        \node[roundednode, fill=blue] (ForwardDiff) at (1.5, -2.8) {\texttt{ForwardDiff.jl}};
        \node[roundednode, fill=blue] (Enzyme) at (3, -2) {\texttt{Enzyme.jl}};

        \node[roundnode, fill=purple] (DI) at (1.5, -1) {\texttt{DI.jl}};

        \draw[edge] (Optim) to (DI);
        \draw[edge] (DiffEq) to (DI);
        \draw[edge] (Turing) to (DI);

        \draw[edge] (DI) to (Zygote);
        \draw[edge] (DI) to (Enzyme);
        \draw[edge] (DI) to (ForwardDiff);
      \end{tikzpicture}
    }
    \subcaption{Julia (now)}
  \end{subfigure}
  \caption{Comparison of the AD ecosystems in Python and Julia for applications to optimization, differential equations and probabilistic programming}
  \label{fig:ecosystems}
\end{figure}

To facilitate comparison, the missing ingredient was a common API compatible with every AD package in Julia.
Hence we developed \texttt{DifferentiationInterface.jl}\footnote{\url{https://github.com/JuliaDiff/DifferentiationInterface.jl}}, or \texttt{DI} for short, a unified frontend to a dozen AD backends (see Appendix \ref{sec:list} for a list).
It acts as a single entry point for AD users, abstracting away implementation details to focus on the desired output.
Such a common interface offers flexibility and modularity, without compromising on speed.
The benefits of this approach are synthesized on Figure \ref{fig:ecosystems}, where we show that for common machine learning tasks, the number of necessary bindings decreases drastically.

\section{Design principles}

The main concepts underpinning \texttt{DI} are backends, operators and preparation.
While the first two were inspired by a previous proof of concept \citep{schaferAbstractDifferentiationjlBackendAgnosticDifferentiable2022}, we argue that preparation is the key novelty unlocking peak performance and widespread adoption.

\paragraph{Backends.}
A backend is a Julia object which represents a choice of AD package, combined with package-specific parameters (e.g.\ the differentiation mode, or the taping behavior).
For historical reasons, backend types are defined in a lightweight third-party package called \texttt{ADTypes.jl}\footnote{\url{https://github.com/SciML/ADTypes.jl}}.
Thanks to Julia's multiple dispatch mechanism, user code is specialized on the provided backend to generate efficient assembly.
A minimal example is displayed in Listing \ref{lst:example} (left), where we compute the gradient of the squared Euclidean norm with \texttt{ForwardDiff.jl} \citep{revelsForwardModeAutomaticDifferentiation2016}.
In this code sample, only two lines need to change for a different AD package to be used instead (see Appendix \ref{sec:composability} for a more involved example).

\begin{listing}
  \begin{minipage}{0.42\textwidth}
\begin{minted}{julia}
using DifferentiationInterface
# change 2 lines to switch backends
import ForwardDiff
back = AutoForwardDiff()

f(x) = sum(abs2, x)
x = float.(1:5)
g = gradient(f, back, x)
\end{minted}
  \end{minipage}
  \hfill
  \begin{minipage}{0.56\textwidth}
\begin{minted}{julia}
gradient!(f, g, back, x)  # in-place
y, g = value_and_gradient(f, back, x)  # primal
\end{minted}
\begin{minted}{julia}
prep = prepare_gradient(f, back, zero(x))
for iteration in 1:1000
    x -= 0.1 * gradient(f, prep, back, x)  # fast
end
\end{minted}
  \end{minipage}
  \caption{Example uses of \texttt{DI}}
  \label{lst:example}
\end{listing}

\paragraph{Operators.}
\texttt{DI} provides a set of 8 differentiation operators.
First-order operators are \inline{pushforward} (Jacobian-vector product), \inline{pullback} (vector-Jacobian product), \inline{derivative}, \inline{gradient}, and \inline{jacobian}.
Second-order operators are \inline{second_derivative}, \inline{hvp} (Hessian-vector product), and \inline{hessian}.
Each operator possesses 4 variants: in-place and out-of-place, with or without primal output, as demonstrated in Listing \ref{lst:example} (top right).
If an AD package does not natively provide a given operator, \texttt{DI} takes over with a default chain of fallbacks until reaching the lowest-level operators \inline{pushforward} and \inline{pullback}.
For instance, (1) \inline{gradient} relies on \inline{pullback}, (2) \inline{hvp} combines \inline{pushforward} with \inline{gradient} and (3) \inline{hessian} calls \inline{hvp} once per input dimension.

\paragraph{Preparation.}
AD is very useful inside iterative procedures like gradient descent, where the same function is differentiated many times on different inputs.
In such settings, it is worth preparing for repeated differentiation, paying a one-time cost to speed up each subsequent execution (sometimes by orders of magnitude, see Appendix \ref{sec:preparation}).
This initial step takes different forms depending on the AD package.
For some, preparation will record an execution tape or perform the source transformation.
For others, it will precompute useful data like basis vectors, or preallocate caches, or even perform symbolic simplification.

\texttt{DI}'s most valuable contribution consists in hiding this complexity from users, who are only asked to provide a typical input with the correct type and size (e.g.\ \enquote{I want to prepare the gradient of my function for double-precision vectors of length 10}).
Then, whatever information or memory the AD package needs is encapsulated in the result of preparation, and can be reused as many times as necessary.
The corresponding syntax is demonstrated in Listing \ref{lst:example} (bottom right).
By encouraging \texttt{DI} users to adopt this preparation mechanism across their computationally intensive tasks, we set the stage for arbitrarily complex AD methods to be used transparently.
In particular, this is crucial for the efficient application of sparse AD techniques (see below).

\section{Supported features}

Not only is it easier to call each package with \texttt{DI}, the interface also delivers additional abilities.
Here we list a few features which are natively supported by a subset of AD packages, but which \texttt{DI} makes easily accessible for all (or most) of them.

\paragraph{Contexts.}
When the function to differentiate takes several arguments, usually not all derivatives are needed.
\texttt{DI} assumes that only the first argument is actually differentiated (a common limitation of AD packages in Julia), but it supports additional non-differentiated arguments of two types: \inline{Constant}s and \inline{Cache}s.
The first kind is dedicated to fixed parameters, whereas the second kind is for buffer storage which gets overwritten by the function to avoid new allocations.

\paragraph{Sparsity.}
\texttt{DI} allows the efficient computation of sparse Jacobian and Hessian matrices, using techniques reviewed in \citet{gebremedhinWhatColorYour2005}.
This functionality is made possible by two additional packages: \texttt{SparseConnectivityTracer.jl}\footnote{\url{https://github.com/adrhill/SparseConnectivityTracer.jl}} for sparsity pattern detection, and \texttt{SparseMatrixColorings.jl}\footnote{\url{https://github.com/gdalle/SparseMatrixColorings.jl}} for coloring problems.
Both of these preliminary steps happen during the preparation phase, so that their high cost is amortized by subsequent computations.
See \citet{hillSparserBetterFaster2025} for details and benchmarks.

\paragraph{Backend combination.}
A single differentiation mode is not always enough.
For instance, HVPs and Hessian matrices are most efficiently computed in forward-over-reverse mode \citep{dagreouHowComputeHessianvector2024}.
When individual AD packages do not include both modes, \texttt{DI} supports the creation of a \inline{SecondOrder} object to stack different backends and achieve the required behavior.
Similarly, the \inline{MixedMode} wrapper is used for sparse Jacobians that combine forward and reverse passes.

\paragraph{Backend translation.}
Some functions are differentiable with one AD backend but not the other.
In such cases, writing custom rules for every package is tedious.
Instead, \texttt{DI} includes utilities for translating between packages, essentially saying \enquote{when package A tries to differentiate this function, use package B under the hood}.
This is implemented as a wrapper \inline{DifferentiateWith(f, other_backend)}.

\paragraph{Testing.}
The main appeal of having access to several AD packages is to figure out which one is best for a given application.
To make this easier, \texttt{DI} comes with a sibling package \texttt{DifferentiationInterfaceTest.jl} (or \texttt{DIT}), which contains testing and benchmarking functionality.
The user only has to define a testing scenario, and they can quickly compare correctness and speed between the candidate backends, with standardized result reporting (see Appendix \ref{sec:benchmark} for a demonstration).

\section{Perspectives}

So far, \texttt{DI} has mostly been developed for use cases in scientific machine learning, where problems have a moderate dimension and parallelism is hard to achieve due to numerous scalar operations.
A natural prospect would be to improve GPU support and testing, which is crucial for large-scale workloads.

Another avenue for extension is backend-agnostic definition of custom rules.
\texttt{DI} is a unfied way to call AD, but it does not yet include rule-building utilities, which vary much more between backends.
A first common interface was already introduced by \texttt{ChainRules.jl} \citep{whiteJuliaDiffChainRulesjlV17232025}, but the advent of mutation-friendly AD systems like \texttt{Enzyme.jl} \citep{mosesInsteadRewritingForeign2020,mosesReversemodeAutomaticDifferentiation2021} brings new challenges for standardization.

\clearpage


\acks{
  Adrian Hill acknowledges funding from the German Federal Ministry of Education and Research under the grant BIFOLD25B.
  
  The authors want to thank the people who have inspired \texttt{DI}, contributed to it or tried it out themselves. In alphabetical order, those include:
  Fredrik Bagge Carlson,
  Valentin Churavy,
  Jadon Clugston,
  Vaibhav Dixit,
  Francis Gagnon,
  Hong Ge,
  Patrick Mogensen,
  Alexis Montoison,
  William S. Moses,
  Avik Pal,
  Qingyu Qu,
  Chris Rackauckas,
  Frank Schäfer,
  Mohamed Tarek,
  Will Tebbutt,
  Frames Catherine White,
  Penelope Yong,
  and many others.
}

\bibliography{DI}

\clearpage

\appendix

\section{Supported AD packages} \label{sec:list}

Table \ref{tab:list} lists every AD package that \texttt{DI} provides an interface to.
Together, these cover a large majority of AD use cases in Julia (see \citet{sapienzaDifferentiableProgrammingDifferential2024} for a recent review of the ecosystem).
The taxonomy of paradigms is taken from \citet{margossianReviewAutomaticDifferentiation2019}.

\begin{table}[H]
  \centering
  \begin{tabular}{@{}llll@{}}
    \toprule
    \textbf{Package}                                                                               & \textbf{Paradigm}        & \textbf{Modes} & \textbf{Reference}                                      \\ \midrule
    \href{https://github.com/JuliaDiff/ChainRulesCore.jl}{\texttt{ChainRules.jl}}                  & Source transformation    & Both           & \cite{whiteJuliaDiffChainRulesjlV17232025}                                                        \\
    \href{https://github.com/JuliaDiff/Diffractor.jl}{\texttt{Diffractor.jl}}                      & Source transformation    & Forward        &                                                         \\
    \href{https://github.com/EnzymeAD/Enzyme.jl}{\texttt{Enzyme.jl}}                               & Source transformation    & Both           & \makecell[l]{\cite{mosesInsteadRewritingForeign2020}    \\\cite{mosesReversemodeAutomaticDifferentiation2021}} \\
    \href{https://github.com/brianguenter/FastDifferentiation.jl}{\texttt{FastDifferentiation.jl}} & Symbolic differentiation & -              &                                                         \\
    \href{https://github.com/JuliaDiff/FiniteDiff.jl}{\texttt{FiniteDiff.jl}}                      & Finite differences       & -              &                                                         \\
    \href{https://github.com/JuliaDiff/FiniteDifferences.jl}{\texttt{FiniteDifferences.jl}}        & Finite differences       & -              &                                                         \\
    \href{https://github.com/JuliaDiff/ForwardDiff.jl}{\texttt{ForwardDiff.jl}}                    & Operator overloading     & Forward        & \cite{revelsForwardModeAutomaticDifferentiation2016}    \\
    \href{https://github.com/bmad-sim/GTPSA.jl}{\texttt{GTPSA.jl}}                                 & Operator overloading     & Forward        &                                                         \\
    \href{https://github.com/chalk-lab/Mooncake.jl}{\texttt{Mooncake.jl}}                         & Source transformation    & Reverse        &                                                         \\
    \href{https://github.com/JuliaDiff/PolyesterForwardDiff.jl}{\texttt{PolyesterForwardDiff.jl}}  & Operator overloading     & Forward        & \cite{mesterDifferentialMethodsAssessing2022}           \\
    \href{https://github.com/JuliaDiff/ReverseDiff.jl}{\texttt{ReverseDiff.jl}}                    & Operator overloading     & Reverse        &                                                         \\
    \href{https://github.com/JuliaSymbolics/Symbolics.jl}{\texttt{Symbolics.jl}}                   & Symbolic differentiation & -              & \cite{gowdaHighperformanceSymbolicnumericsMultiple2022} \\
    \href{https://github.com/FluxML/Tracker.jl}{\texttt{Tracker.jl}}                               & Operator overloading     & Reverse        &                                                         \\
    \href{https://github.com/FluxML/Zygote.jl}{\texttt{Zygote.jl}}                                 & Source transformation    & Reverse        & \makecell[l]{\cite{innesDontUnrollAdjoint2019}          \\ \cite{innesDifferentiableProgrammingSystem2019}}\\ \bottomrule
  \end{tabular}
  \caption{List of AD packages supported by \texttt{DI}}
  \label{tab:list}
\end{table} 

\section{More examples}

The code examples displayed below were run on a 2023 MacBook Pro M3 using Julia v1.11.5.
The API corresponds to the versions \texttt{DI} v0.6.52 and \texttt{DIT} v0.9.6, it may change in future releases (following semantic versioning).

\subsection{Composability} \label{sec:composability}

In Listing \ref{lst:composability}, we demonstrate the ease of switching the backend object inside a Hessian computation.
First, we use a simple forward-mode backend, which is slow in high dimension (top frame).
Second, we improve performance by switching to forward-over-reverse mode with \texttt{SecondOrder} (middle frame).
Third, we further speed things up by leveraging sparsity with \texttt{AutoSparse} (bottom frame).
This modularity allows the user to experiment and find the best AD option in a given scenario, without diving into the documentation of each package.

\begin{listing}
\begin{minted}{julia}
using DifferentiationInterface, Chairmarks
import ForwardDiff, ReverseDiff
using SparseArrays, SparseConnectivityTracer, SparseMatrixColorings

forward = AutoForwardDiff()
reverse = AutoReverseDiff(; compile=true)
sparsity_detector = TracerSparsityDetector()
coloring_algorithm = GreedyColoringAlgorithm()

f(x) = sum(diff(x) .^ 2)
xsmall, xbig, x0big = float.(1:5), float.(1:1000), zeros(1000)
\end{minted}
  \begin{minipage}[t]{0.47\textwidth}
\begin{minted}{julia}
back1 = forward

p1 = prepare_hessian(f, back1, x0big)







\end{minted}
  \end{minipage}
  \hfill
  \begin{minipage}[t]{0.50\textwidth}
\begin{minted}{julia-repl}
julia> hessian(f, back1, xsmall)
5×5 Matrix{Float64}:
  2.0  -2.0   0.0   0.0   0.0
 -2.0   4.0  -2.0   0.0   0.0
  0.0  -2.0   4.0  -2.0   0.0
  0.0   0.0  -2.0   4.0  -2.0
  0.0   0.0   0.0  -2.0   2.0

julia> @b hessian($f, $p1, $back1, $xbig)
5.463 s (42620 allocs: 18.106 GiB, ...)
\end{minted}
  \end{minipage}

  \medskip

  \begin{minipage}[t]{0.47\textwidth}
\begin{minted}{julia}
back2 = SecondOrder(forward, reverse)

p2 = prepare_hessian(f, back2, x0big)







\end{minted}
  \end{minipage}
  \hfill
  \begin{minipage}[t]{0.50\textwidth}
\begin{minted}{julia-repl}
julia> hessian(f, back2, xsmall)
5×5 Matrix{Float64}:
  2.0  -2.0   0.0   0.0   0.0
 -2.0   4.0  -2.0   0.0   0.0
  0.0  -2.0   4.0  -2.0   0.0
  0.0   0.0  -2.0   4.0  -2.0
  0.0   0.0   0.0  -2.0   2.0

julia> @b hessian($f, $p2, $back2, $xbig)
91.667 ms (3780 allocs: 354.360 MiB, ...)
\end{minted}
  \end{minipage}

  \medskip

  \begin{minipage}[t]{0.47\textwidth}
\begin{minted}{julia}
back3 = AutoSparse(
    SecondOrder(forward, reverse);
    sparsity_detector,
    coloring_algorithm
)

p3 = prepare_hessian(f, back3, x0big)




\end{minted}
  \end{minipage}
  \hfill
  \begin{minipage}[t]{0.50\textwidth}
\begin{minted}{julia-repl}
julia> hessian(f, back3, xsmall)
5×5 SparseMatrixCSC{Float64, Int64}
with 13 stored entries:
  2.0  -2.0    ⋅     ⋅     ⋅ 
 -2.0   4.0  -2.0    ⋅     ⋅ 
   ⋅   -2.0   4.0  -2.0    ⋅ 
   ⋅     ⋅   -2.0   4.0  -2.0
   ⋅     ⋅     ⋅   -2.0   2.0

julia> @b hessian($f, $p3, $back3, $xbig)
116.417 μs (9 allocs: 55.188 KiB)
\end{minted}
  \end{minipage}
  \caption{Switching backends within \texttt{DI}}
  \label{lst:composability}
\end{listing}

\subsection{Benchmarking} \label{sec:benchmark}

In Listing \ref{lst:benchmark}, we showcase the functionalities of \texttt{DIT} for comparing perfomance across various backends.
Users are free to define arbitrary scenarios on which to test differentiation operators.
Here, we measure the time it takes to compute the gradient of the squared Euclidean norm, for inputs of increasing dimension.
In addition to benchmarking, \texttt{DIT} provides similar utilities to check derivative correctness against a reference value.

\subsection{Preparation impact} \label{sec:preparation}

\begin{figure}
  \centering
  \includegraphics[width=\textwidth]{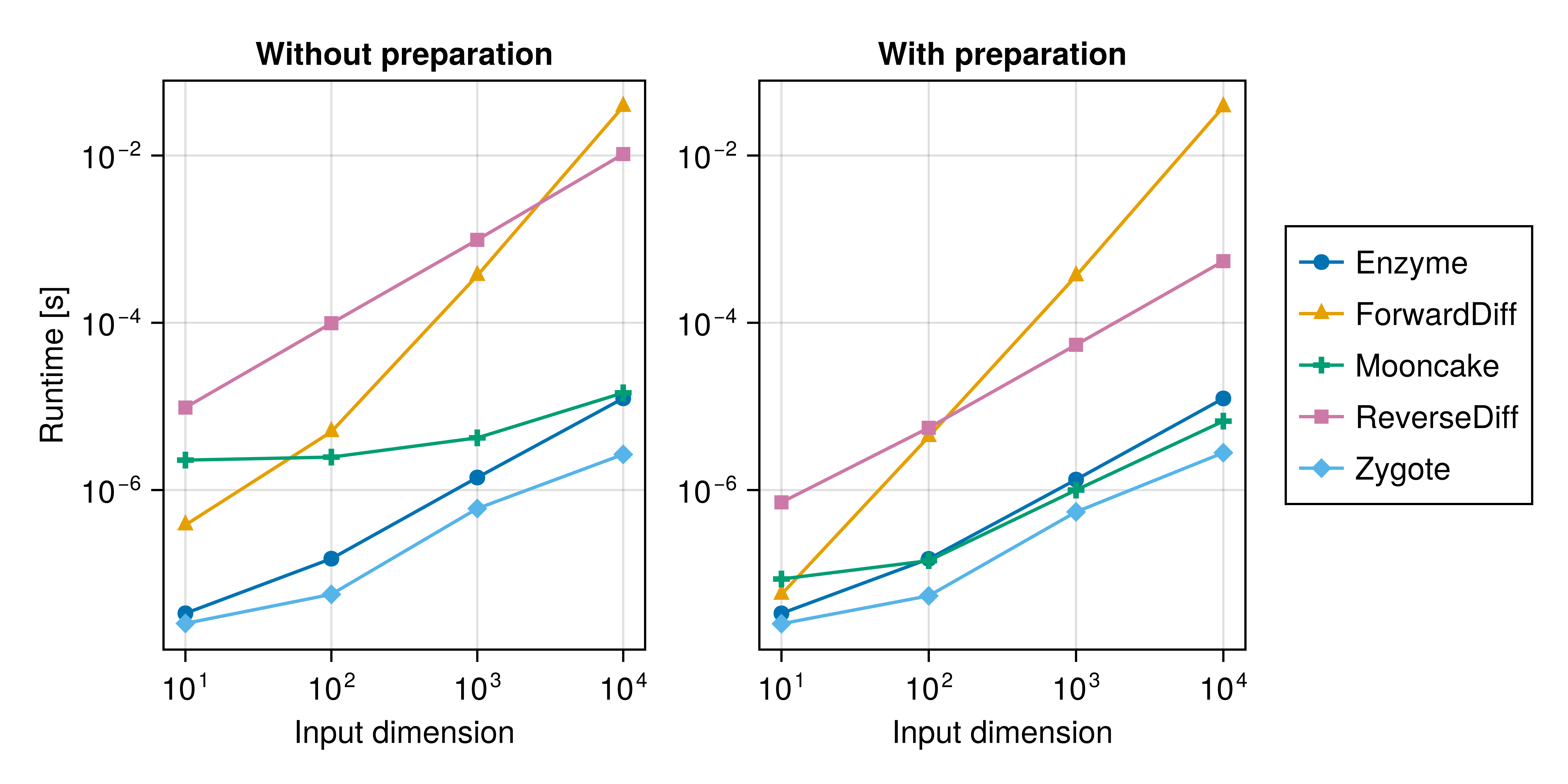}
  \caption{Impact of preparation on gradient performance of $f : x \mapsto \lVert x \rVert^2$}
  \label{fig:benchmark}
\end{figure}

The runtime measurements generated by Listing \ref{lst:benchmark} are displayed in Figure \ref{fig:benchmark}, and they show that the influence of preparation is very backend-dependent.
Of course, it also depends on the test function and the differentiation operator.
In this specific case, \texttt{Enzyme.jl} and \texttt{Zygote.jl} do not benefit from preparation at all.
Meanwhile, the other three backends are greatly sped up by preparation, due to the way they work internally.
\texttt{ForwardDiff.jl} uses preparation to pre-allocate the necessary memory for repeated forward passes, while \texttt{Mooncake.jl} performs source transformation and \texttt{ReverseDiff.jl} compiles an execution tape.
Note that the user does not need to know any of these implementation details to exploit the full abilities of these packages, or to compare them.

\begin{listing}
\begin{minted}{julia}
using DifferentiationInterface, DifferentiationInterfaceTest
import Enzyme, ForwardDiff, Mooncake, ReverseDiff, Zygote

f(x) = sum(abs2, x)
x_candidates = [rand(10), rand(100), rand(1_000), rand(10_000)]

scenarios = [Scenario{:gradient,:out}(f, x; res1=2x) for x in x_candidates]

backends = [
    AutoEnzyme(; mode=Enzyme.Reverse),
    AutoForwardDiff(),
    AutoMooncake(; config=nothing),
    AutoReverseDiff(; compile=true),
    AutoZygote(),
]

df = benchmark_differentiation(backends, scenarios; benchmark=:full)
\end{minted}
\caption{Comparing backends with \texttt{DIT}}
\label{lst:benchmark}
\end{listing}

\end{document}